\documentclass[10pt,twocolumn,twoside]{arxiv}

\usepackage{aas_macros} 

\usepackage{csquotes}

\usepackage{abstract} 
\usepackage{titling}
\usepackage{authblk}

\usepackage{amsmath}
\usepackage{xspace}
\usepackage[capitalize]{cleveref}
\Crefname{equation}{Equation}{Equations}
\crefname{equation}{Eq.}{Eqs.}
\Crefname{figure}{Figure}{Figures}
\crefname{figure}{Fig.}{Figs.}
\Crefname{table}{Table}{Tables}
\crefname{table}{Tab.}{Tabs.}
\Crefname{section}{Section}{Sections}
\crefname{section}{Sec.}{Secs.}

\DeclareMathOperator*{\argmin}{arg\,min}

\title{Rediscovering orbital mechanics with machine learning}

\author[1,2]{Pablo Lemos \thanks{p.lemos@sussex.ac.uk}}
\author[3,2]{Niall Jeffrey \thanks{n.jeffrey@ucl.ac.uk}} 
\author[4]{Miles Cranmer}
\author[4,5,6,7]{Shirley Ho} 
\author[8]{Peter Battaglia}

\affil[1]{Department of Physics  and Astronomy, University of Sussex,Brighton, BN1 9QH, UK}
\affil[2]{University College London, Gower St, London, UK }
\affil[3]{Laboratoire de Physique de l'Ecole Normale Sup\'erieure, ENS, Universit\'e PSL, CNRS, Sorbonne Universit\'e Universit\'e de Paris, 
Paris, France}
\affil[4]{Department of Astrophysical Sciences, Princeton University, Princeton, New Jersey 08544, USA}
\affil[5]{Center for Computational Astrophysics, Flatiron Institute, New York, NY 10010, USA}
\affil[7]{Department of Physics, New York University, New York, NY 10011, USA}
\affil[6]{Department of Physics, Carnegie Mellon University, Pittsburgh, PA 15217, USA}
\affil[8]{DeepMind, London, N1C 4AG, UK}

\leadauthor{Lemos} 

\date{}

\renewcommand{\maketitlehookd}{%
\begin{abstract}
\noindent
We present an approach for using machine learning to automatically discover the governing equations and hidden properties of real physical systems from observations. 
We train a ``graph neural network''
to simulate the dynamics of our solar system's Sun, planets, and large moons from 30 years of trajectory data. We then use symbolic regression to discover an analytical expression for the force law implicitly learned by the neural network, which our results showed is equivalent to Newton's law of gravitation.
The key assumptions that were required were
 translational and rotational equivariance, and Newton's second and third laws of motion. 
 Our approach correctly discovered the form of the symbolic force law. 
Furthermore, our approach did not require any assumptions about the masses of planets and moons or physical constants. They, too, were accurately inferred through our methods.
Though, of course, the classical law of gravitation has been known since Isaac Newton, our result serves as a validation that our method can discover unknown laws and hidden properties from observed data. More broadly this work represents a key step toward realizing the potential of machine learning for accelerating scientific discovery.
\end{abstract}
}

\newcommand\pysr{\textit{PySR}\xspace}
\newcommand\eureqa{\textit{eureqa}\xspace}

\begin{document}

\maketitle

\begin{quote}
\noindent
\textit{
Discover the force of the skies O Men: once recognised it can be put to use.
} - Johannes Kepler
\end{quote}

Machine learning (ML) has prompted dramatic advances in many scientific disciplines, most commonly through its ability to process large, complex sets of observations and make predictions about desired properties. 
From particle physics~\cite{Bourilkov:2019yoi} to structural biology~\cite{jumper2021} to cosmology~\cite{He:2018ggn},
ML methods help find patterns in large data sets \cite{2020arXiv200304480R, Arpaia:2020mvq}, classify different objects \cite{2021MNRAS.502..206S}, and perform parameter inference~\cite{Nolan:2020zjo, Green:2020dnx, deep_learning_des}, as well as more groundbreaking applications such as autoregressive language models~\citep{2020arXiv200514165B} and predicting protein structure~\citep{Jumper2021HighlyAP}, and protein function prediction \citep{Gligorijevic786236} 
However, there have been relatively few applications of ML to one of the most fundamental parts of science: theory discovery. 
Here, we demonstrate a new approach for using real data and established scientific frameworks to discover both 
physical laws and unobserved properties of a complex physical domain---our solar system. 
In this work, we use observations of the orbital trajectories of the Sun, planets, and moons to re-discover Newton's law of gravitation, as well as the masses of these bodies.
This process is analogous to the process followed by scientists when they develop scientific theories 
and constrain parameters from observations. Usually, a scientist proposes theories through mathematical formulae and evaluate them against data, and our approach described here automates key components of this endeavor.

Our approach
involves two stages: training a learned simulator on observed data, then performing symbolic regression on components of the simulator which correspond to physical laws.
In~\cite{2019arXiv190905862C,Cranmer:2020wew} we described an initial version of our general approach, applied to simulated data, but here we have introduced several innovations, incorporated a new technique for simultaneously inferring unobserved properties of the system (e.g., the masses of the bodies), and, most importantly, applied it to real data and showed it can recover correct physical laws.
Applying our approach to real data presented new challenges: the data are noisy, and their dynamic range spans several orders of magnitude; the dataset is partial (we only provide 31 objects; leaving out other massive bodies); and, crucially, the masses of the bodies are not observed, and therefore needed to be discovered at the same time.

The first stage's learned simulator is based on graph networks (GN)~\cite{battaglia2018relational}, which are deep neural networks that can be trained to approximate complex functions on graphs. Here the (relative) positions and velocities of the solar system's sun, planets, and moons are represented as nodes of the input graph, and possible physical interactions (e.g., forces) between the bodies are represented by the graph's edges. GN-based simulators have been trained to accurately model N-body and more complex particle- and mesh-based systems in recent years~\cite{2016arXiv161200222B,sanchez2020learning,pfaff2020learning}, though they have never been trained on real observations until now. We fit the GN-based simulator to 30 years of observed solar system trajectories, where the training procedure optimized the parameters of the GN's neural network ``edge function'', which plays the role of computing forces~\citep[see][]{Cranmer:2020wew,2019arXiv190905862C}.

In the second stage, we isolate this edge function, and apply symbolic regression to fit an analytical formula to it. Our best fitting expression was the correct Newtonian formula for gravitational force. We then re-fit the unobserved (relative) masses of the bodies using our discovered equation, and found a nearly perfect fit to the true masses. We could then simulate the solar system dynamics using the discovered equation and re-learned masses, and get a very close correspondence to the true observed trajectories. 

The reason we adopt this two-step approach, instead of applying symbolic regression directly on the data, is that symbolic regression is not practical or efficient. Because the learned simulator is a neural network, it is differentiable, and thus fitting it to the real data is very efficient. By contrast, the symbolic regression procedure involves an expensive search using evolutionary algorithms, which would take orders of magnitude longer. The differentiability is also effective for fitting continuous quantities, such as the masses of the bodies. So by fitting a neural network simulator first, then applying symbolic regression to only that component of the learned simulator we were interested in, we reduced the cost of the equation discovery a great deal, and make the problem tractable for our symbolic regression code. 

There are several reasons to prefer a symbolic expression, instead of settling with a learned simulator. Naturally, describing physical phenomena with compact symbolic formulations supports scientific interpretation, and can interface with existing symbolically defined physical theories. By contrast, the knowledge stored within a trained neural network cannot easily interface with existing theories---how can one interpret the thousands, millions, or even billions of weights within a neural network, or communicate that knowledge effectively to others? Beyond interpretability, the symbolic expression we extracted was more accurate than the predictions of the neural network, due to the strong bias toward simplicity in the symbolic regression. We found it could generate far more accurate predictions than the GN-based simulator, and by virtue of that fact that it is correct, it should generalize to any scale, while the neural network is only accurate when the statistics of the input match what it was trained on. In other words, the analytical expression can model galactic dynamics, while the learned simulator cannot.

Symbolic regression, also known as automated equation discovery, has been explored for decades in the context of scientific discovery, for example~\citet{Langley1987}'s BACON~\cite{Langley1977BACONAP}, COPER~\citep{kokar}, FAHRENHEIT/EF~\citep{LANGLEY1989283, Zembowicz} and LAGRANGE~\citep{CoTodorovski1997DeclarativeBI}. 
More recent work~\cite{bongard,Schmidt81} introduced the symbolic
regression package \eureqa, which has been applied to finding symbolic formulae for Lagrangians, Hamiltonians, repeated
sub-equations, etc., without relying on known constants or strong priors on the physical nature of the system.
Though there have been many advances in search 
techniques~\citep[e.g.][]{eql, grammarvae, Udrescu:2019mnk, brunton, koopman1,koopman2, deepymod,atkinson,universalode,pidl,fluidpdealg, bayesianmachinescientist, 10.1162/evco_a_00278, Brunton3932, 2019arXiv190402107C}, in this work we use the neural network-symbolic regression technique we first introduced in \cite{Cranmer:2020wew}, which extends symbolic regression to high-dimensional input such
as graphs by using a neural network as an intermediate stepping-stone model. We have also released an open-source symbolic regression software library, which we used in this work~\pysr\footnote{\url{https://github.com/MilesCranmer/PySR}} \cite{pysr, cranmer2020discovering} .

\begin{figure}[t]
\centering
\includegraphics[width=.99\linewidth]{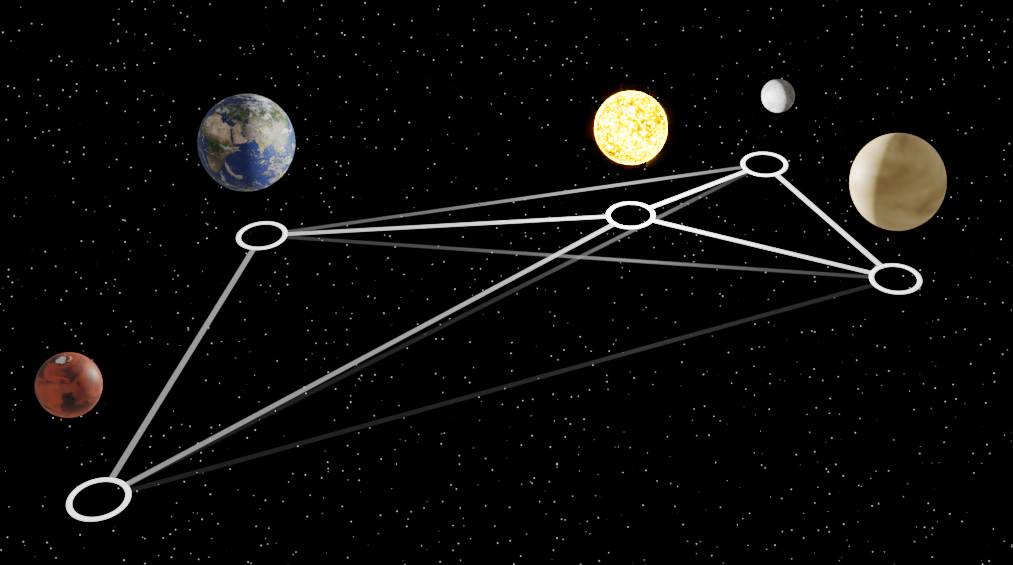}
\caption{Schematic rendering of the Sun, Mercury, Venus, Earth, and Mars, with the corresponding graph structure our learned simulator uses. The graph's nodes represent the bodies, and the brightness of the edges is proportional to the strength of the gravitational interaction between them. A video version of this figure will be made available upon publication.
}
\label{fig:graph}
\end{figure}

It is important to emphasize that there is no way to \enquote{discover} new theories without imposing \textit{some} constraints, inductive biases, or other assumptions on the process. For example, mathematical axioms are required to define quantitative theories; the concepts of space and time are required to specify equations of motion; and a physical mechanics formalism, such as classical mechanics, is required to define specific dynamical laws, such as Hooke's law or the Hamiltonian of a many-body system. Here, our approach leverages the fact that an N-body system can be represented as a graph; and that these systems are translationally equivariant~\citep{2016arXiv160207576C}. Our learned simulator incorporates Newton's laws of motion in that the learned scalar for each node is multiplicative in scaling the model's output to acceleration; and finally, our equation search prioritizes expressions which are simple, which is analogous to Occam's razor.

Ultimately we believe our approach should be viewed as a tool which can help scientists make parts of their discovery process more efficient and systematic, rather than as a replacement for the rich domain knowledge, scientific methodology, and intuition which are essential to scientific discovery.

\section*{Model}
\label{sec:methods}
Our two-step approach first fits a GN-based learned simulator to model the observed trajectories, then uses symbolic regression to fit analytical formulae to internal components of the learned simulator, which we designed to have direct correspondences to classical mechanics' force law. Within our learned simulator, we used one trainable scalar value per body, which scaled the predicted acceleration for the body, and thus can be interpreted as a mass (i.e., the equation \textit{acceleration} = (\textit{GN output}) $/$ (\textit{scalar} corresponds to $F=ma$).

\subsection*{Graph network-based learned simulator}
The input to our GN-based learned simulator, $g_\theta$, is a graph, $(V, E)$, which represents the physical system, where the set of $N^v$ bodies are represented as nodes, $V=\{v_i\}_{i=1:N^v}$, and relationships between pairs of bodies are represented as directed edges, $E=\{(s_k, r_k, \mathbf{e}_k)\}_{k=1:N^e}$. Each $v_i$ node attribute contains a trainable scalar variable that is fixed across all input graphs, and is analogous to mass as described below. 
The $s_k$ and $r_k$ edge attributes are integers which index the sender and receiver nodes, respectively. The $\mathbf{e}_k$ edge attribute is the spatial displacement vector between the two corresponding bodies.
Because we assume we do not know which bodies interact, we instantiate edges from each body to every other body, which allows us to model all possible pairwise interactions. 

To simulate the bodies' dynamics, the model predicts the per-body accelerations, $a_i$, by explicitly imposing Newton's second and third laws of motion. The GN contains an \enquote{edge function}, $\mathbf{e}'_k=f_\mathrm{GN}(v_{r_k}, v_{s_k}, \mathbf{e}_k; \theta)$, with trainable parameters $\theta$, which computes an interaction vector, $\mathbf{e}'_k$, along each edge, which is analogous to a force. For the two directed edges between a pair of bodies, $(i, j, \mathbf{e}_k)$ and $(j, i, \mathbf{e}_l)$, rather than computing distinct $\mathbf{e}'_k$ and $\mathbf{e}'_l$ we instead compute just one and set the other equal to its negative, $\mathbf{e}'_l = - \mathbf{e}'_k$, in accordance with Newton's third law's \enquote{equal and opposite} principle. Next, for each body, $i$, all of its incoming interaction vectors are summed, $\bar{\mathbf{e}}'_i = \sum_{\{k \vert r_k = i\}} \mathbf{e}'_k$, analogous to superposition of forces to compute net force. Finally, the per-node output accelerations, $\hat{a}_i = \bar{\mathbf{e}}'_i / v_i$ 
are computed by dividing each node's pooled interactions by the scalar node attribute, $v_i$, which, following Newton's second law's $F=m a$, gives $v_i$ the semantics of \enquote{mass} and $\bar{\mathbf{e}}'_i$ the semantics of \enquote{net force}\footnote{Note, in practice we use $\log(v_i)$, in order to reduce the dynamic range of $v_i$.}. The sun's scalar attribute is fixed to $1$ to fix the degeneracy of scale between the learnable GN and learnable scale. The details of the neural networks are described in the Experimental Methods below.

Our learned simulator $g_\theta$ 
is trained by supervised learning, where the discrepancies between the model's predicted accelerations and the true observed accelerations are minimized with respect to the trainable model parameters using gradient descent, 
\begin{equation}
\theta^*, V^*=\argmin_{\theta, V} \mathbb{E}_{(E, A) \sim \mathcal{D}_{\mathrm{train}}} \ell_\mathrm{GN}(g(V, E;\theta), A)
\qquad ,
\end{equation}
where $A$ are the true observed accelerations associated with some input $(V, E)$, $\ell_\mathrm{GN}$ is an error metric, and $\mathcal{D}_{\mathrm{train}}$ is the empirical distribution which represents the observed system states (represented by the relative displacements between bodies, $E$) and accelerations used for training. While the edge attributes, $E$, vary as the positions of the bodies in the system change, the scalar per-node attributes, $V$, are trainable variables which are constant across inputs. By minimizing the error with respect to $V$, we are fitting the masses for each body in the system, which we will compare to the known masses of the solar system bodies in the Results

\subsection*{Symbolic regression of force function}
Once the learned simulator was trained, we performed symbolic regression to fit an explicit symbolic formula to the GN-based force function. 
We created a dataset of force function inputs, $(v_{r_k}, v_{s_k}, \mathbf{e}_k) \in \mathcal{D}_{\mathrm{SR}}$, and used the symbolic regression procedure to search for an expression, $f_\mathrm{SR}$, which minimizes,
\begin{equation}
f^*_\mathrm{SR}=\argmin_{f_\mathrm{SR}} \mathbb{E}_{x \sim \mathcal{D}_{\mathrm{SR}}} \ell_\mathrm{SR}(f_\mathrm{SR}(x), f_\mathrm{GN}(x; \theta^*))
\qquad ,
\end{equation}
where $x=(v_{r_k}, v_{s_k}, \mathbf{e}_k)$ sampled from the empirical symbolic regression training distribution, $\mathcal{D}_{\mathrm{train}}$, and $\ell_\mathrm{SR}$ is an error metric.

The symbolic regression procedure explores a space of analytic expressions and selects one or more which predict the target, $f_\mathrm{GN}(v_{r_k}, v_{s_k}, \mathbf{e}_k; \theta^*)$, accurately, while also minimizing the complexity of the discovered expression.
The space of symbolic expressions is large due to the combinatorial number of ways the operators, variables, and constants can be composed (e.g., if there are $M$ possible discrete symbols, then there are $M^L$ possible symbol strings of length $L$, but actually the constants are effectively real-valued rather than discrete).
Because it is fundamentally a discrete problem, we cannot compute gradients or perform gradient descent, as with the GN-based simulator's training.

\begin{figure*}[t]
    \centering
	\includegraphics[width=0.99\linewidth]{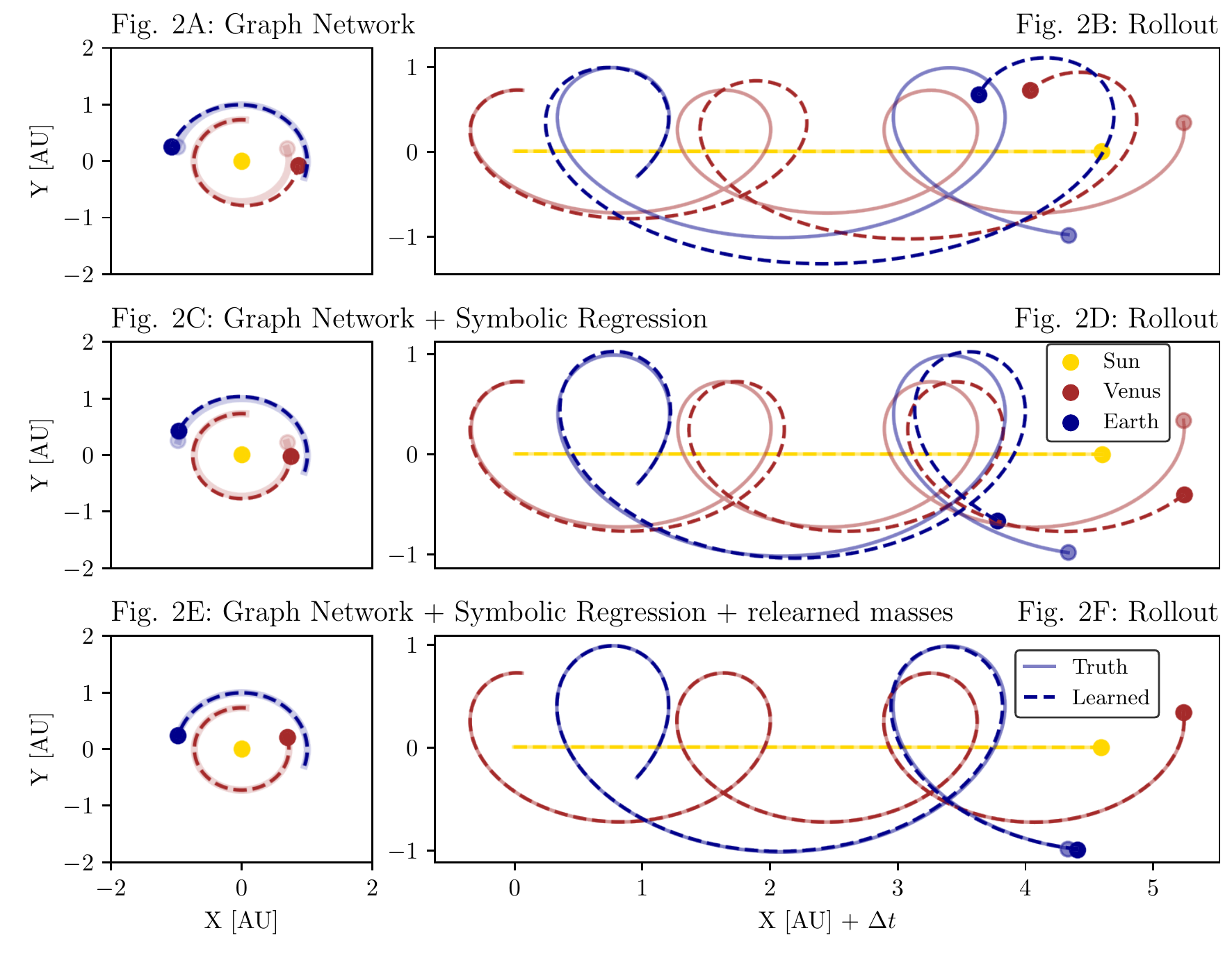} 
    \caption{\label{fig:rollout}
    Comparison between data (continuous line) and integrated prediction (dashed line) of same bodies evolved from same initial conditions, for the interactions predicted by the learned simulator (top, 2A and 2B), the symbolic regression expression (center, 2C and 2D) and the same expression, after re-estimating the masses (bottom, 2E and 2F). 
    The left panels show the orbits for six months. The right panels show the same orbits but for 21 months, with a time displacement along the x-axis, which allows us to visualize the trajectory for a longer time period.
} 
\end{figure*}

\section*{Experimental methods}
\label{sec:experiments}

\subsection*{Data}
We use solar system data from NASA's HORIZONS On-Line Ephemerys 
System\footnote{\url{http://ssd.jpl.nasa.gov/?horizons}}~\citep{1996DPS....28.2504G, 2001DPS....33.5813G}.
We extract orbits for 31 bodies: the Sun, all planets, and those moons 
which have a mass above $10^{18} \ \mathrm{kg}$. Whilst more bodies could have been considered, we expect their gravitational influence to be small, therefore we do not expect that their omission will affect our results. 
We use data from January 1980 to January 2013 with a time step of $30$ minutes, and use the first $30$ 
years of data (approximately one full orbit of Saturn) for training, and the last three for validation. From 
the HORIZONS interface, we extract positions and velocities in Cartesian coordinates, with the solar system barycenter as the reference frame.

From this data, we extract the pair-wise displacement vectors between bodies and each body's acceleration vector (calculated from changes in the velocities) at every step. 
Relative displacements serve as the input to our model, meaning that our model is equivariant to a translated reference frame. The accelerations serve as the truth for our model training.

Therefore, Our input graph has $N^v=31$ nodes; each node with one trainable scalar, and a single edge connecting every pair of bodies $N^e = N^v (N^{v}-1)/2 = 465$; each containing three coordinates giving the distances between bodies along each spatial axis.

\subsection*{Graph network implementation details}
The GN uses a TensorFlow~\citep{tensorflow2015-whitepaper} model with three-layer multilayer perceptrons (MLPs) and 128 hidden nodes per layer. The model also contains the trainable scalar properties of the nodes $v_i$, which are backpropagated simultaneously with the weights of the neural network. 
Furthermore, our model has the following properties:

\begin{itemize}
\item {\bf Activation function}: We use a hyperbolic tangent (``tanh'') as the activation function in our networks. While this is slower than the 
very commonly used Rectified Linear Unit (ReLU) activation function~\citep{2018arXiv180308375A}, our problem is very susceptible
to the dying ReLU problem~\citep{2019arXiv190306733L} due to the very different values of both inputs and outputs.
\item {\bf Loss function}: For the loss function, we use the relative mean weighted error: 

\begin{equation}
\label{eq:loss} 
{\rm Loss} = \sum \frac{\left( A -g(V, E;\theta) \right)^2 }{ A^2}.
\end{equation}
The reason we use the relative mean weighted error is again due to the large dynamic ranges experienced in our dataset, so that every body is emphasized equally during training, not only the ones with large accelerations.

\item {\bf Spherical coordinates}: Our GN takes as inputs a 3-vector for every pair of bodies representing the displacements, 
and outputs a second 3-vector which corresponds to the force 
However, due to the large dynamic range of input displacements, 
we transform the input displacement from Cartesian into spherical coordinates, using $\log_{10}$ to transform the magnitude, as inputs. Similarly, the output force is assumed to be in spherical coordinates, whose magnitude component is transformed through an exponential function back into Cartesian. This allows the GN to learn forces of very different magnitudes, without requiring the parameter distribution inside the GN to have a large dynamic range itself.

\item {\bf Data augmentation}: a random three-dimensional rotation is applied to the input graph at every training iteration. This serves as data
augmentation, useful for our limited-size training data. It also prevents biases from being created inside the model, and encourages a learned rotational equivariance: for example, the solar system is largely confined to a plane (which could bias along the rotational axis), and some planets are not observed to complete an entire orbit in the training set (which could bias in their particular direction).

\item {\bf Training noise}: During training, we corrupted the input states with Gaussian noise to improve the model's robustness to error over long rollouts at test time. This technique has been used widely in GN-based learned simulators~\cite{sanchez2020learning,pfaff2020learning}: it is believed to help the model close the gap between the distribution of training input states, which are always from the true observations, and rollout input states, which are predicted by the model and incur some error.

\item {\bf Early stopping}: We stop training once a threshold was reached where 20 epochs experienced no improvement in the validation loss, to prevent overfitting.

\item {\bf Multiple runs}: To estimate the uncertainty in our estimation of the masses, we 
repeat the minimization procedure for ten different random seeds, and calculate a mean and standard 
deviation in the mass estimates from the different best fits.

\item {\bf Local minima}: Our loss function has multiple local minima, in which gradient descent is at risk 
of becoming ``stuck''. Therefore, we restart the minimization when training stops with a validation loss larger than 0.5 (for reference, the best fit loss is typically close to 0.05). 

\end{itemize}

\subsection*{Symbolic regression implementation details}
We used the \pysr~\citep{pysr,Cranmer:2020wew} library\footnote{\url{https://github.com/MilesCranmer/pysr}} for symbolic regression which was developed by some of the authors. \pysr~is an open-source analog to \eureqa{}~\citep{Schmidt81}, which has a Python API and also supports distributed computation and custom operators and losses.

\pysr\ uses a tree search algorithm to produce a set of candidate equations, that go from some input features (in our cases displacements and learned masses) to some outputs (in our case forces). The objective of \pysr\ in this case is to find a `simple' and interpretable equation that resembles the interaction predicted by the learned simulator. We do this because we are looking for physical laws that can explain nature with simple equations, as opposed to the high complexity of a neural network. To accomplish this target simplicity, \pysr\ assigns to each proposed equation a score, calculated as the ratio between the increase in accuracy (in our case, the decrease in our error metric $\ell_\mathrm{SR}$) and the increase in complexity with respect to the previous proposed equation. The complexity is calculated from the number of terms and operators that are used in the equation. More details about this can be found in a coming paper. It is clear that different options could be use for both the complexity, accuracy, and score calculations. Therefore, we do not claim that \pysr\ uniquely obtains the perfect equation. Instead, its role is to produce a set of candidate equations from the infinite set of possible ones, with a complexity that is orders of magnitude lower than the complexity of a deep neural network.

For this application, we select a dataset $\mathcal{D}_{\mathrm{SR}}$ consisting of 500 data points that were not used during training of the learned simulator. Each of these points contains as inputs the learned scalar variables and displacements between a pair of randomly selected bodies at a random time step $x = (v_{r_k}, v_{s_k}, \mathbf{e}_k)$, and as outputs the corresponding interaction learned by the GN $f_\mathrm{GN} (x)$. We add the norm of the displacement vector $\left| \mathbf{e}_k \right|$ as an extra input. 
The allowed operators between these input quantities are addition, subtraction, multiplication and division. The maximum complexity allowed for the equations is 40. We use as our constant optimizer the Broyden–Fletcher–Goldfarb–Shanno (BFGS) algorithm~\citep{10.1093/imamat/6.1.76, 10.1093/comjnl/13.3.317, Goldfarb1970AFO, Shanno1970ConditioningOQ} with 10 iterations, and our error metric $\ell_\mathrm{SR}$ is a MSE loss function between the GN interaction $f_\mathrm{GN} (x)$ and the proposed equation $f_\mathrm{SR} (x, \theta^*)$.

\section*{Results}
\label{sec:results}

\begin{figure}[t]
    \centering
	\includegraphics[width=0.99\linewidth]{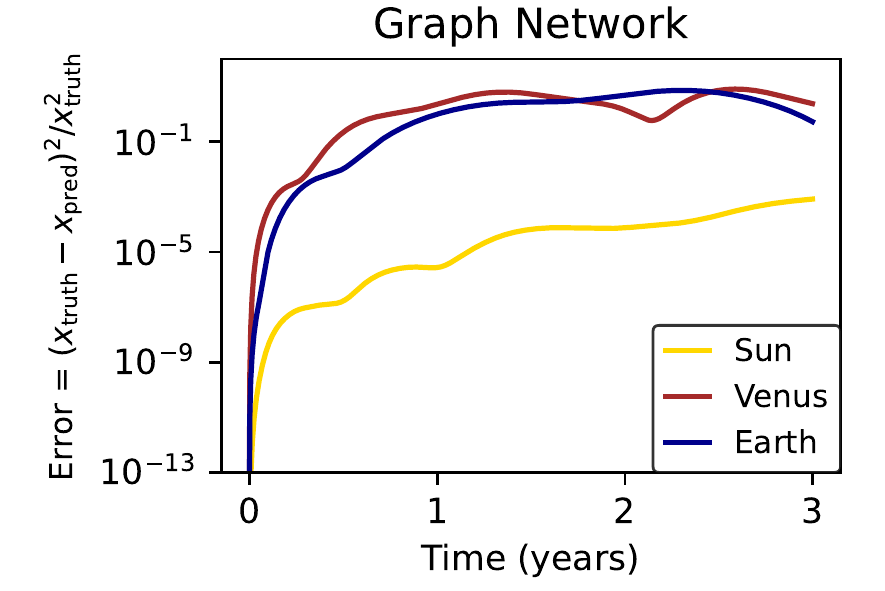} 
    \caption{\label{fig:losses}
    The loss per step, calculated using~\cref{eq:loss}, from integrating the trajectories for the bodies shown in~\cref{fig:rollout} using the learned simulator. A version of this figure for all other bodies can be found in the appendix.
} 
\end{figure}

\subsection*{Learned Simulator Performance}
Our model learns to predict interactions between bodies which generally agree with the observed 
accelerations. 
The predicted next-step relative accelerations, $(x_{\rm truth} - x_{\rm pred})^2 / x_{\rm truth}^2$, have error of around $0.2\%$ on the validation data, averaged over all bodies and time steps. These accelerations can be time-integrated to roll out predicted trajectories, as 
shown for the trajectories of the Sun, Venus, and Earth
in the top panel of ~\cref{fig:rollout} (2A and 2B).
The predicted dynamics agree with the ground truth observations over short time intervals, and begin to deviate after several months. This is not surprising because the model is trained to predict only the next time step, 30 minutes in the future, and the strong non-linearity of N-body dynamics lead to small errors rapidly growing over the rollout. 
Figure~\cref{fig:losses} shows how the rollout error accumulates over the 3 years of validation data for the same bodies used in~\cref{fig:rollout}. Similar figures for all other bodies are available in the appendix.

This shows that 
the GN-based simulator can learn dynamics from real data, rather than simulated data as in previous work. Because our focus here was on symbolic discovery, the learned simulator we used was relatively simple compared to recent GN-based models~\citep{pfaff2020learning,sanchez2020learning}, but with more powerful methods we expect the accuracy would be even greater.

\begin{figure*}[h]
    \centering
	\includegraphics[width=0.99\linewidth]{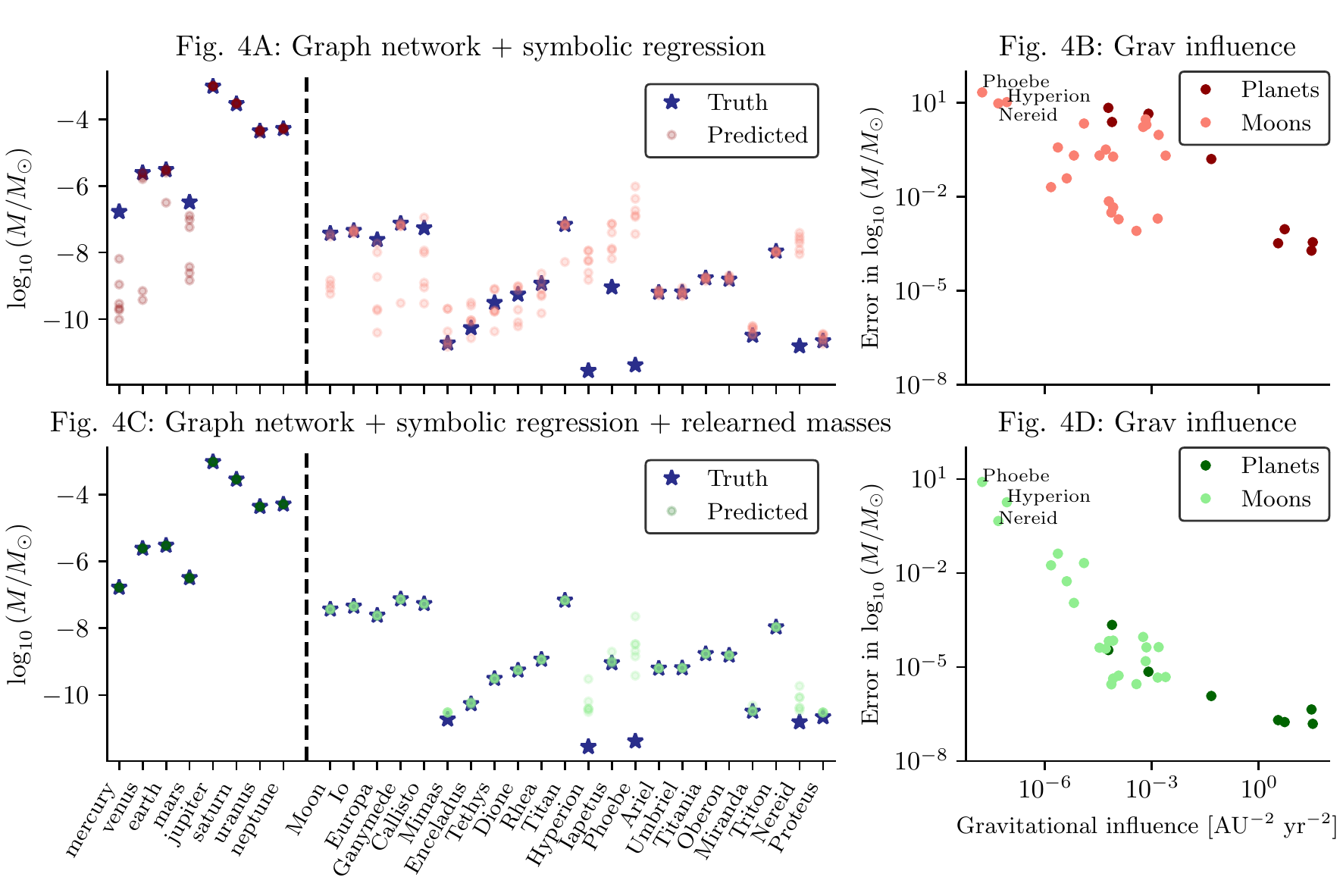} 
    \caption{\label{fig:masses} {\it Left}: Comparison of the learned scalars $v_k$ relative to the Sun $v_0$ and the known logarithmic masses of the solar system bodies (in units of solar mass).
    The multiple red points represent the different mass values for each random seed, and are an attempt to capture the model's uncertainty. The red dashed line separate planets and satellites). {\it Right}: Negative correlation between the gravitational influence exerted by each body, and the error in its mass estimate. The plot shows clearly how the masses of the bodies that have negligible gravitational influence on others cannot be estimated by our two-step algorithm. The top two panels use the masses trained with the learned simulator, while the bottom two use the masses re-estimated using the symbolic expression.
    } 
\end{figure*}

\subsection*{Hidden Property Inference}

The learned scalar properties $v_i$, which scale the predicted accelerations and thus play the role of mass, are shown in the top left part of~\cref{fig:masses} (4A) along with the
masses per body. The multiple plotted values for each body represent the fit masses from different training runs with different random initializations (discarding 
runs that get stuck in local minima as described in the previous section), and help give a sense of the uncertainty in the estimates. The results indicate that the scalar quantities learned by our model roughly match the true masses for the bodies represented in our dataset with a mean percent error $\sim 9.1 \%$ calculated over all bodies and all different initializations.

The errors between our model's inferred masses and the true ones demonstrate an interesting pattern: bodies which have little effect on other bodies' accelerations tend to have higher mass error.
We computed the gravitational influence of a body $n$ as the sum of gravitational potentials experienced by all other bodies that result from body $n$,
\begin{equation}
    {\rm grav. \ influence}_n = \sum_{i \neq n} V_{\rm grav}(i,n) = \sum_{i \neq n} - {G M_n \over \left| \vec{x}_n - \vec{x}_i \right|}, \label{eq:grav-influence}
\end{equation}
which sums over all bodies except for $n$, and where $V_{\rm grav}$ is the gravitational potential and $G$ is the gravitational constant.
We calculate this gravitational influence for each body, as the mean of the gravitational influence summed over time, to account for the fact that bodies with eccentric orbits might have a larger gravitational influence at certain points as they get closer to other bodies. 

The top right panel of~\cref{fig:masses} (4B)
plots the error in the estimate of the mass $(1/N_{\rm inits}) \sum_{\rm inits} (\log_{10} M_{\rm true} - \log_{10} M_{\rm pred})^2 $, where the sum is over all 10 random initializations, as a function of gravitational influence in Equation~\ref{eq:grav-influence}.
The figure shows a clear negative correlation (Pearson correlation coefficient of $-0.64$ in log space) between the error in the mass estimate and the gravitational influence. In other words, bodies that have a strong influence on the rest have very accurate masses, while those that are not very influential have poor mass estimates. 
For example, Mercury and Venus do not have moons, and Mars' moons were too small to be included in our dataset, and thus they do not have nearby bodies to affect. Similarly, the moons Phoebe, Hyperion, and Nereid have small masses and thus have little influence on their planet and nearby moons. Thus the mass errors are to be expected: the ``equivalence principle''\footnote{Because $F=m_{\text{body}} a$,
and the $F$ function includes $m_{\text{body}}$ in the numerator, $m_{\text{body}}$ cancels and is not required to compute its own acceleration.} 
holds that for bodies which impart negligible gravitational influence on other bodies in the system, and thus do not influence other bodies' accelerations, their masses are, in ML parlance, ``unidentifiable'', meaning such masses are difficult to estimate accurately. 

\begin{figure*}
	\includegraphics[width=0.99\linewidth]{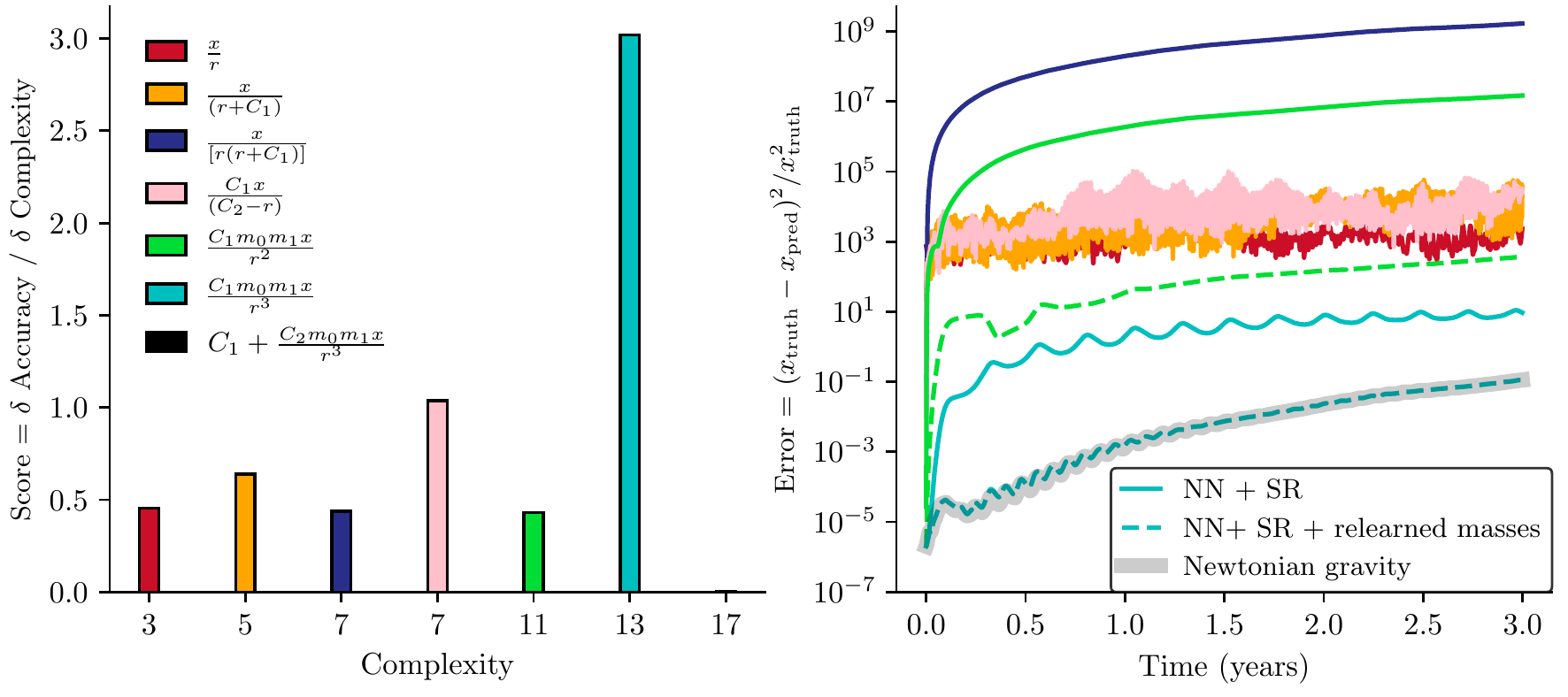} 
    \caption{\label{fig:sr} 
    {\it Left}: Discovered equations from our learned simulator across different random initializations. 
    The different equations are sorted in order of increasing complexity along the x-axis. 
    The output variable in all cases is $F_x$, but 
    similarly results are obtained for $F_y$ and $F_z$.
    The y-axis shows the equation score, which balances loss and complexity, as described in~\cite{pysr}.
    {\it Right}: Loss per step on predicted orbits using each equation, summed over all planets. Note that we do not show the complexity 17 equation, as the results perfectly overlap with those of the complexity 13 equation. For the two equations that have masses, we plot the error using the masses from the learned simulator (continuous line) and from refitting the masses after obtaining the equation (dashed). The grey, thicker line is the error obtained using Newtonian gravity with the correct masses.
    } 
\end{figure*}

\subsection*{Symbolic Discovery}

Our symbolic regression procedure correctly discovered Newton's law of gravity. The left panel in Figure~\cref{fig:sr} shows candidate equations obtained by \pysr. The top 6 equations were those with the highest score, of the over one hundred million tested, sorted in order of increasing complexity. The highest bar corresponds to one with the same form as Newton's gravity. The seventh, rightmost bar, was the best equation which had higher complexity than the best equation, which we plotted in order to demonstrate that increasing complexity does not necessarily provide a better score.

To show how each equation fared in predicting orbital trajectories, the right panel of Figure~\ref{fig:sr} plots their respective rollout errors over the three years of validation data, and compares with the true data. The lowest error is the cyan line's equation, which corresponds to Newton's law of gravity,
\begin{equation}
\label{eq:newton}
\vec{F} = - \frac{G_{\rm learned} M_1 M_2 }{ r^3}\vec{r},
\end{equation}
where $M_i$ are the masses learned by our learned simulator, shown in the top panel of~\cref{fig:masses}.

Our symbolic regression method also learns a value of the gravitational constant which is very similar to the true one.
Note, similar to how we fit masses relative to the Sun's mass, constants which include a mass unit are also relative to some reference mass. 

We plotted the rollout trajectories over the three years of validation data using the best-fit equation in the second row of~\cref{fig:rollout} (2C and 2D).
The orbits predicted by the discovered symbolic expression are more accurate over time than those from the learned simulator. This means that despite \pysr\'s fitted formula being simpler than the learned simulator's neural network-based one, it yields more accurate predictions.

\subsection*{Relearning the masses}
Having determined the correct form of the interactions between bodies, we can then \textit{re-}estimate the hidden properties. We replaced the MLP edge function within the learned simulator's GN with the best-fit symbolic expression~(\cref{eq:newton}), and re-trained the mass and gravitational variables in the same manner as we originally trained the learned simulator.
The bottom-left panel of~\cref{fig:masses} (4C) shows
how the mass estimates are far more accurate than they were from the original learned simulator training, with a mean percent error calculated over all bodies and iterations of $\sim~1.6\%$, more than a factor of five lower than before relearning the masses. Similarly, the bottom-right panel of~\cref{fig:masses} (4D) clearly shows how the negative correlation between error in mass estimate and gravitational influence in other bodies is greatly strengthened by re-learning the masses, with the Pearson correlation coefficient in log-space going from $-0.64$ (before re-learning the masses) to $-0.87$ (after re-learning). 

We also predicted new trajectories with the symbolic equation and re-learned masses (bottom panel of~\cref{fig:rollout}, 2E and 2F), which shows the trajectories are far more accurate than those obtained directly from both the learned simulator, and the symbolic equation with original masses. 
The blue dashed curve in the right part of~\cref{fig:sr} shows the difference between data and prediction for these re-estimated masses. The figure clearly shows how this outperforms the learned simulator and symbolic regression, and perform just as well as Newtonian gravity using the correct parameters (thick grey curve).
Therefore, our algorithm obtains the correct equation for Newtonian gravity (\cref{fig:sr}) and very accurate values for the masses of the bodies (bottom part of \cref{fig:masses}), using only the orbit data, graph structure, and some inductive biases as input information.

\section*{Discussion \& Conclusions}
\label{sec:discussion}

Our results show that our two-step approach---training a neural network simulator with physical inductive biases, then interpreting what it has learned using symbolic regression---is a powerful tool for discovering physical laws from real observations. We (re-)discovered Newton's formula for gravitational force from observed trajectories of the Sun, planets, and moons of our solar system, and made accurate estimates of hidden properties of the system.

While our method allows us to re-discover Newton's formula and the masses, it is important to note that this was only possible through the use of inductive biases, particularly Newton's second and third law, and spherical symmetry. Furthermore, we made use of choices such as spherical coordinates and logarithmic units, which facilitated the learning. This illustrates that while automated theory formation with machine learning is possible, it does require some prior knowledge. Our understanding of the system can therefore greatly facilitate the task of discovering physical laws with machine learning.  

While automated theory formation is a very promising and exciting field of work, it is important 
to consider the limitations of this procedure. First, while we can provide a rough estimate of the
uncertainty in the mass estimates by running with multiple random seeds, this does not
produce a true estimate of the errors, and instead shows multiple local minima where the algorithm terminates. To perform Bayesian inference on the mass estimates, we would need to model the posterior distribution on each mass, which cannot be done with our current 
graph network algorithm, which uses gradient descent to produce point estimates. Bayesian 
neural networks could provide a future avenue for this. 
Second, while our algorithm can provide 
a scientist with candidate equations that produce a good fit to the data, as shown in~\cref{fig:sr}, the specific scoring function used to measure the quality of equations (e.g., complexity vs. accuracy) warrants further exploration. The scientist's preferences for what make a ``good'' equation should be expressed, and more generally, the candidate equations should be viewed as a narrower palette of choices, which should be subject to further experimentation.

This work offers a new way of marrying modern machine learning methods with automatic theory formation, and demonstrates its efficacy in the context of complex real-world data. Even though the law we discovered is already known of course, the purpose of this work is to confirm that known laws and hidden properties are \textit{discoverable} with our method. This is a key step toward building more sophisticated tools for automating the process of scientific discovery, in particular data-driven theory formation and evaluation.

\acknow{PL acknowledges STFC Consolidated Grants ST/R000476/1 and ST/T000473/1. We also thank the organisers of the University College London CDT DIS seminars, where the collaboration that led to this work begun.}

\showacknow{} 

\bibliography{refs}

\appendix

\section*{Appendix}
\label{sec:appendix}
\begin{figure*}
\centering
\includegraphics{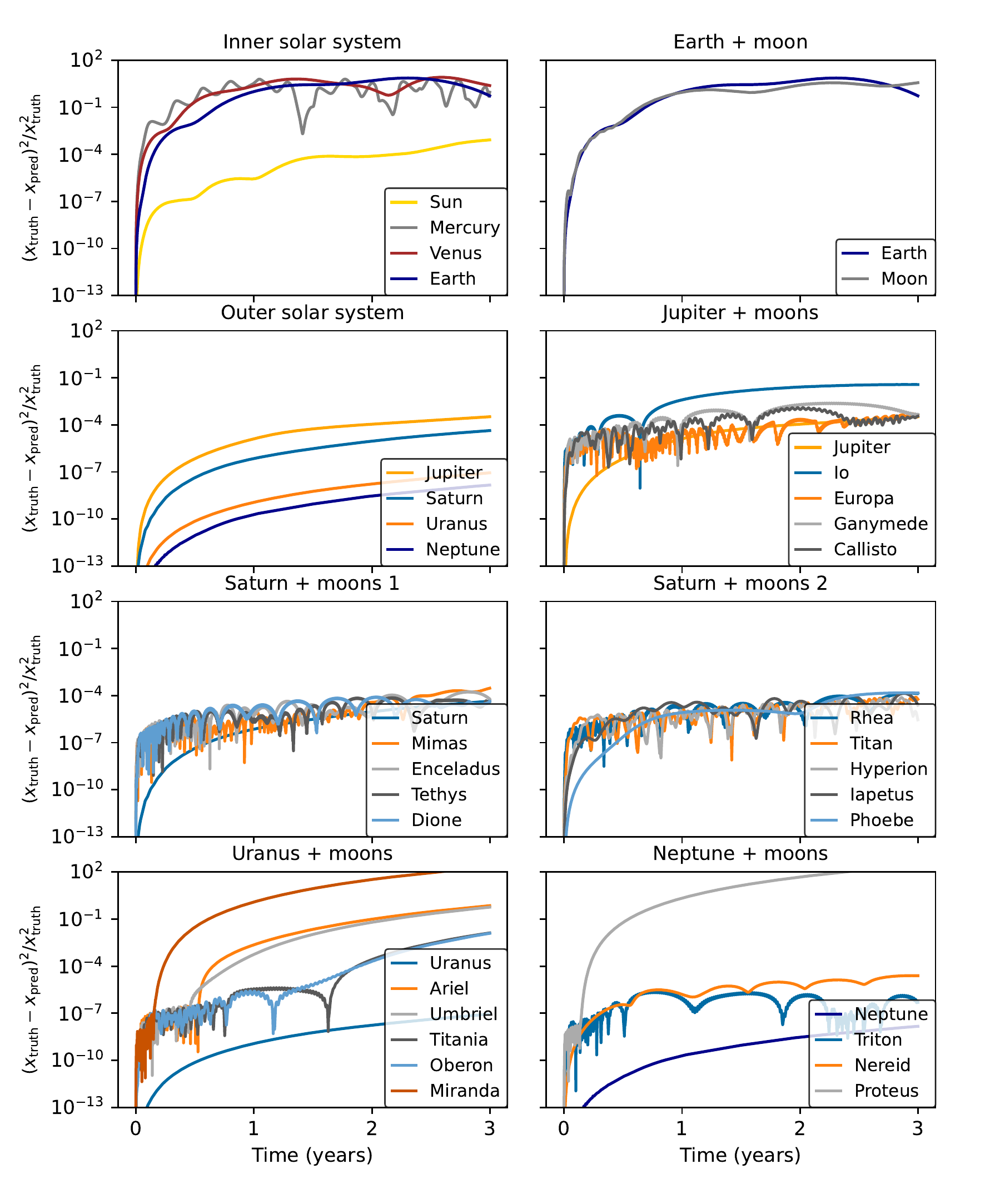}
\caption{ The loss per step, calculated using~\cref{eq:loss}, from integrating the trajectories for the bodies shown in~\cref{fig:rollout} using the learned simulator.
\label{fig:rollout_errors}}
\end{figure*}

\cref{fig:rollout_errors} shows the rollout errors for all the bodies used in this paper, defined as $(x_{\rm truth} - x_{\rm pred})^2 / x_{\rm truth}^2$ where $x_{\rm pred}$ is the trajectory obtained when integrating using the interaction learned by the graph network, $and x_{\rm truth}$ is the real data. These curves are calculated from data that was not used during training. This plot is similar to Fig. 3 in the text, but showing all bodies in our system. 

\end{document}